%% file: AASKAII_TDEs_R2.tex
\documentclass[a4paper,11pt]{article}
\usepackage{aaskaiid}
\usepackage{orcidlink}
\usepackage{booktabs}
\providecommand{\farcs}{.\!\!^{\prime\prime}}

\newcommand{\AAstar}{AA$^{\ast}$}
\newcommand{\termdefs}{Terminology used throughout this chapter is defined: SKA--Low is the 50--350\,MHz aperture-array telescope in Western Australia; SKA--Mid is the 0.35--15.4\,GHz dish array in South Africa, including MeerKAT; AA4 denotes the design-baseline array of 512 SKA--Low stations and 197 SKA--Mid dishes; \AAstar\ denotes the staged-delivery array of 307 SKA--Low stations and 144 SKA--Mid dishes. A \textit{commensal slow-transient survey} is an image-plane transient search run on data taken by another primary survey, with revisits on day-to-week timescales. }

\title{Tidal Disruption Events with the SKA}
\ShortTitle{TDEs with the SKA}

\author[1,\dagger]{T. An\orcidlink{0000-0003-4341-0029}}
\ShortName{An et al.}
\author[2]{A. J. Goodwin\orcidlink{0000-0003-3441-8299}}
\author[2,\dagger]{J. C. A. Miller-Jones\orcidlink{0000-0003-3124-2814}}
\author[3]{M. P\'erez-Torres\orcidlink{0000-0001-5654-0266}}
\author[4]{L. Rhodes\orcidlink{0000-0003-2705-4941}}
\author[5]{X. Shu\orcidlink{0000-0002-7020-4290}}
\author[]{the Transient Science Working Group}

\affiliation[1]{Shanghai Astronomical Observatory, Chinese Academy of Sciences, 80 Nandan Road, Shanghai 200030, China}
\emailAdd{antao@shao.ac.cn}
\affiliation[2]{International Centre for Radio Astronomy Research -- Curtin University, GPO Box U1987, Perth, WA 6845, Australia}
\emailAdd{adelle.goodwin@curtin.edu.au}
\emailAdd{james.miller-jones@curtin.edu.au}
\affiliation[3]{Instituto de Astrof\'\i sica de Andaluc\'\i a, Consejo Superior de Investigaciones Cient\'\i ficas (CSIC),
Glorieta de la Astronom\'\i a S/N, 18008 Granada, Spain}
\emailAdd{torres@iaa.csic.es}
\affiliation[4]{Trottier Space Institute, McGill University, Canada}
\emailAdd{lauren.rhodes@mcgill.ca}
\affiliation[5]{Department of Physics, Anhui Normal University, Wuhu, Anhui, 241002, China}
\emailAdd{xwshu@ahnu.edu.cn}
\affiliation[\dagger]{Chapter co-ordinator}

\abstract{
Tidal disruption events (TDEs) and related nuclear transients probe jet launching, disk formation and circularization, particle acceleration, and the circumnuclear medium (CNM). However, the small fraction of events launching relativistic jets, the weak or delayed radio emission of many thermal TDEs, and the black-hole demographics of galactic nuclei remain poorly understood. SKA, with $\mu$Jy sensitivity, wide bandwidth, long baselines, rapid response, and commensal surveys\footnotemark\ across 50--350\,MHz (SKA--Low) and 0.35--15.4\,GHz (SKA--Mid), will transform this field. Its sensitivity, frequency coverage, and very long baseline interferometry (VLBI) capability will extend radio calorimetry from a few well-studied nearby events to volume-limited samples of non-relativistic outflows. The SKA will bring off-axis and mildly relativistic jets into routine reach, trace CNM density profiles through the evolution of the low-frequency synchrotron turnover, and localize faint off-nuclear transients with sub-arcsecond precision. These data will constrain jet incidence, energetics, geometry, and magnetization, map the CNM through shock interactions, test links to changing-look active galactic nuclei (AGN) and high-energy neutrinos, and identify off-nuclear events associated with recoiling supermassive or intermediate-mass black holes. We discuss the benefits of commensal surveys, rapid triggering, and SKA--VLBI, and provide predictions based on the projected performance of \AAstar\ and AA4. SKA will shift TDE radio studies from detailed case-by-case studies to population-scale inference, advancing studies of jet physics and black-hole demographics.
}

\begin{document}
\include{journal-names}
\maketitle
\footnotetext{\termdefs}

\section{Introduction}

A tidal disruption event (TDE) occurs when a star passes within the tidal radius of a massive black hole and is torn apart; a fraction of the debris returns to pericenter, circularizes, and is accreted, powering luminous emission across the spectrum \citep{Rees1988}. Multi-wavelength observations have revealed two broad classes \citep{Gezari2021}: (i) thermal TDEs, discovered primarily in the optical/UV and soft X-rays, likely dominated by reprocessing and/or disk emission; and (ii) jetted TDEs, emitting hard X-rays and $\gamma$-rays from a relativistic jet launched along our line of sight, with long-lived radio afterglows from external shocks. The field has been punctuated by landmark events. Swift~J1644+57 established the ``jetted TDE'' paradigm, appearing as a bright, variable X-ray source with a powerful radio afterglow \citep{Bloom2011,Burrows2011,Zauderer2011}. ASASSN-14li became the template for nearby thermal TDEs with a well-characterized, non-relativistic radio outflow \citep{Holoien2016,Alexander2016,vanVelzen2016}. More recently, AT2022cmc was the first on-axis jetted TDE discovered optically, demonstrating the feasibility of detecting newly-launched relativistic jets with wide-field optical surveys and following their radio evolution from the earliest times \citep{Andreoni2022,Pasham2023,Rhodes2023}. Together, these events showcase the diverse phenomenology of TDEs in terms of jet power, geometry, circumnuclear medium (CNM) interaction, and timescales.

Radio observations probe the kinetic output of TDE outflows and their coupling to the CNM \citep{Alexander2020,vanVelzen2021}. Synchrotron spectral energy distributions and their time evolution yield shock radii, velocities, magnetic fields, and the energetics of the outflows, while late-time light-curve flattening or rebrightening constrains renewed energy injection, or density gradients, clumps, or shells in the nuclear environment \citep{Chevalier1998,BarniolDuran2013}. For Swift~J1644+57, the long-lived radio afterglow and achromatic breaks have been modeled as a decelerating relativistic jet interacting with the CNM \citep{Zauderer2011,Zauderer2013,Berger2012,Metzger2012}; for ASASSN-14li, multi-epoch radio monitoring indicates a slower, wide-angle outflow propagating through an $r^{-2}$-like density profile \citep{Alexander2016,vanVelzen2016}. For several optically-selected TDEs, including AT2020vwl and AT2018hyz, delayed radio brightening months to years after optical peak suggests evolving outflows and/or changes in the accretion state \citep{Goodwin2025,Cendes2022}.

Despite these advances, major gaps remain. The radio-loud (jetted) fraction appears to be only a few percent, but the selection biases are severe \citep{Alexander2020,vanVelzen2021}. The fraction of ``radio-quiet'' (non-jetted) TDEs with weak or delayed radio outflows is poorly constrained \citep{Goodwin2022,Cendes2022}, and the mapping from SMBH mass/spin and stellar parameters to jet production efficiency remains theoretically and observationally unsettled \citep{Tchekhovskoy2014,BlandfordZnajek1977}. The demographics of low-mass supermassive black holes (SMBHs) and intermediate-mass black holes (IMBHs) inferred from TDEs are still limited by small samples, uncertain selection functions, and incomplete radio follow-up \citep{Wevers2017,Mockler2019,Lin2018}. Moreover, TDE-like nuclear transients (e.g., dramatic changing-look active galactic nucleus (AGN) episodes such as 1ES~1927+654) blur the boundary between TDEs and AGN variability, with radio behavior often unconstrained \citep{Trakhtenbrot2019,Ricci2020}. Finally, the tentative association of high-energy neutrinos with TDEs \citep{Stein2021,Reusch2022}, while intriguing, remains disputed \citep{Cendes2021}, and coordinated radio campaigns will be required to probe hadronic acceleration sites and calorimetry \citep{Mohan2022}.

\begin{figure}
    \centering
    \includegraphics[width=0.8\linewidth]{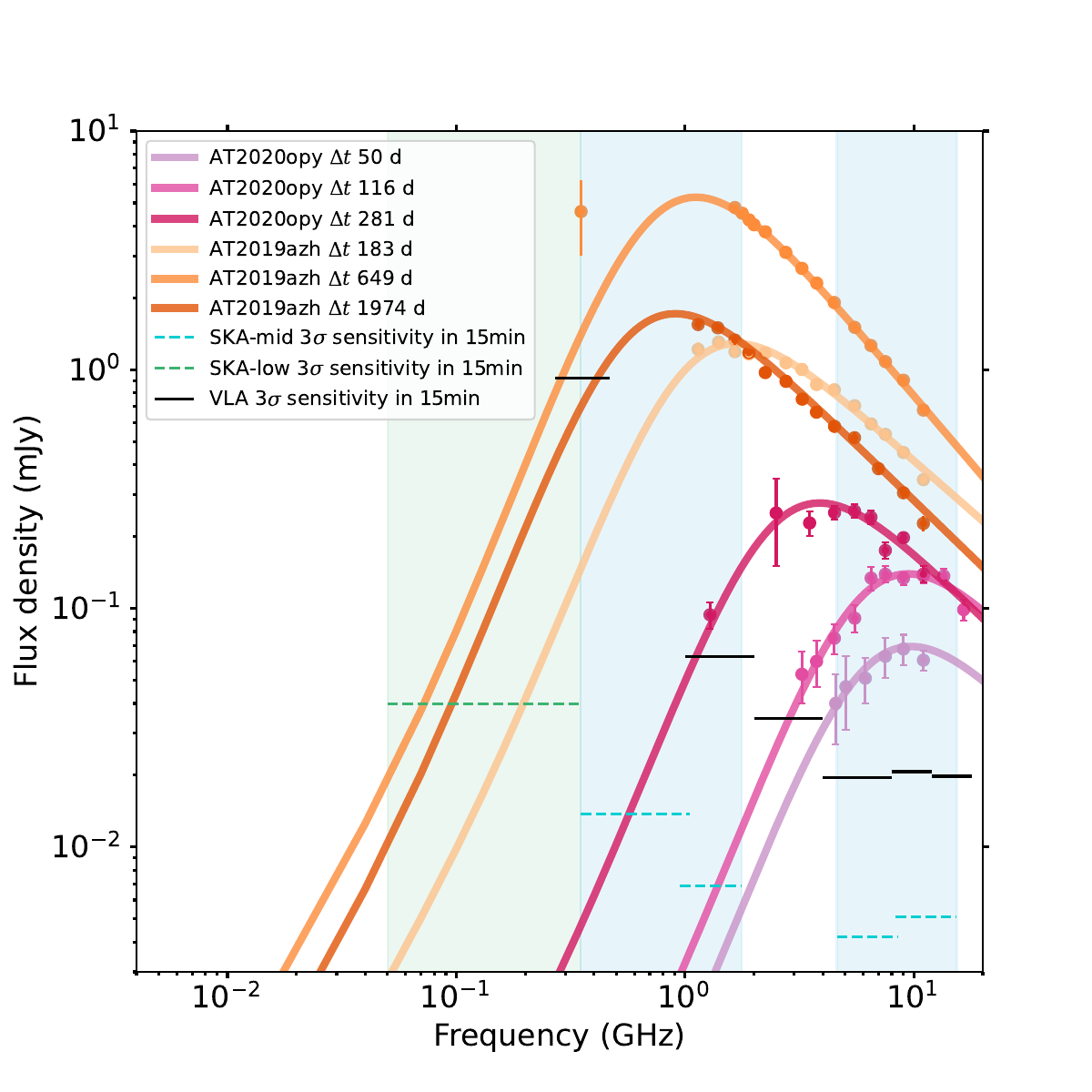}
    \caption{Broadband radio spectra at different times since optical flare of the TDEs AT2019azh \citep[oranges;][]{Goodwin2022,Burn2025} and AT2020opy \citep[pinks;][]{Goodwin2023a}. The sensitivity of the SKA and VLA in various observing bands in a 15\,min observation are indicated with horizontal lines, and the observing bands of the SKA are highlighted in green (low) and blue (mid).  
    AT2020opy is the furthest non-jetted TDE so far observed with broadband radio spectral observations, at a redshift of $z=0.15$. Due to its large distance, the source was only able to be tracked for $\approx200$\,d with the VLA. In contrast, AT2019azh is a relatively nearby (and bright) non-relativistic TDE at a redshift of $z=0.022$ in which broadband radio monitoring has been possible for $>6$\,yr. This figure demonstrates how the SKA will not only allow more distant (faint) TDEs to be detected and characterised, but also will allow the synchrotron spectral peak to be tracked for significantly longer than is currently possible as it evolves to lower frequencies with time.}
    \label{fig:spectra}
\end{figure}

The SKA will provide the high-fidelity data necessary to resolve these long-standing ambiguities in nuclear transient physics. 
SKA--Mid provides a continuum sensitivity of 1.4--4.6\,$\mu$Jy in 15\,min on source across Bands 1 through 5b (Fig.~\ref{fig:spectra}), with baselines up to 150 km enabling sub-arcsecond imaging from 0.95--15.4\,GHz, precisely-measured spectral indices, and high time-resolution variability studies. 
In the same observing time, SKA--Low will provide a sensitivity of 13\,$\mu$Jy\,beam$^{-1}$ at 50--350 MHz, where late-time, optically thin synchrotron emission from expanding TDE shocks may persist for years. Example TDE spectra and nominal improvement on sensitivities of current world-class radio instruments at 0.05--15\,GHz are shown in Figure \ref{fig:spectra}. 
The telescopes will support transient-friendly modes, multiple subarrays, and voltage-beam output for VLBI. Even in the staged \AAstar\ configuration before 2030, these capabilities will bring faint and delayed radio emission from thermal TDEs into reach, tighten constraints on jet geometry (e.g., jet opening angles, Lorentz factors) and energetics, trace CNM structure through the evolution of broadband spectra, and build radio samples large enough for statistically robust population studies.

The SKA will operate within an ecosystem of time-domain surveys and multi-messenger facilities. The Rubin Observatory's Legacy Survey of Space and Time (LSST) will find large numbers of nuclear transients; the Wide Field Survey Telescope (WFST) will help deliver optical classification, and the Cerenkov Telescope Array Observatory (CTAO), and the IceCube Neutrino Observatory extension (IceCube-Gen2) and the Cubic Kilometre Neutrino Telescope (KM3Net) will provide gamma-ray and neutrino triggers, respectively.
The SKA's survey speed, rapid response, and deep targeted follow-up, in synergy with SKA--VLBI \citep{Shu.1.2026.SKA}, are the missing ingredients to decisively map accretion--outflow coupling and black-hole demographics in low-mass galactic nuclei.

In the following sections we outline how the SKA, from its staged \AAstar\ phase to the full AA4 configuration, will quantify TDE rates and energetics, reconstruct accretion histories of dormant SMBHs, and map the physics of their outflows.

\section{Key science questions}

The major science questions that SKA will address in the field of TDEs are outlined below, focusing on jet production, outflow energetics, CNM structure, and black-hole demographics. Some require larger and cleaner samples. Others require access to low radio luminosities, low frequencies, and milliarcsecond localization. A combination of these is largely beyond current facilities.

\subsection{Why do only a few percent of TDEs launch relativistic jets? How are jets connected to the accretion flow?}

Only a small fraction of TDEs produce highly relativistic jets visible as luminous, rapidly variable X-ray emission and long-lived radio afterglows \citep[e.g., Swift~J1644+57, AT2022cmc;][]{Bloom2011,Zauderer2011,Burrows2011,Andreoni2022,Rhodes2023}. Candidate drivers include black hole spin (via the Blandford--Znajek process), the accumulation of large-scale magnetic flux in magnetically arrested disks \citep[MAD;][]{Tchekhovskoy2014}, super-Eddington accretion states \citep{Metzger2016}, and stellar orbital geometry \citep[impact parameter;][]{Guillochon2013}.
However, the mapping from these ingredients to jet launching, Lorentz factor, and opening angle remains unclear. The SKA will address this problem by enlarging the radio-selected TDE sample via its high-sensitivity wide-field radio surveys, capturing both on-axis jetted events and off-axis afterglows, and by measuring early-time variability and spectral evolution to infer Doppler factors and brightness temperatures. AT2022cmc already demonstrated that day-scale radio variability implies large Doppler factors \citep[][]{Yao2024}. Broad frequency coverage from 50\,MHz to 15.4\,GHz will break degeneracies between synchrotron self-absorption and external free--free absorption, tightening constraints on shock microphysics and magnetization.

The few radio-detected on-axis jetted TDEs at $z \lesssim 0.1$ had GHz flux densities of several mJy during the first weeks, and even off-axis or mildly relativistic jets at $z \sim 0.1$--0.3 are expected to produce $\sim$10--100~$\mu$Jy near the synchrotron self-absorption peak, which is detectable with integrations of minutes to hours with the SKA telescopes. High-cadence, multi-frequency SKA monitoring with a few epochs probing hour-to-day timescales during the first $\sim$100\,d can separate scintillation from intrinsic variability and constrain the Doppler ($\delta$) and Lorentz ($\Gamma$) factors from synchrotron spectral modelling. At the population level, combining SKA detection statistics with all-sky optical/X-ray TDE counts yields the true jetted fraction via
$f_{\rm j}\,\approx\, N_{\rm radio}/N_{\rm TDE}$,
where $N_{\rm radio}$ is the number of SKA-detected radio jets (including off-axis detections) and $N_{\rm TDE}$ is the optical/X-ray TDE yield. In the limiting case where only on-axis jets are detected, this reduces to $f_{\rm j}\,\approx\, N_{\rm on}/(N_{\rm TDE} f_{\rm b})$, where $N_{\rm on}$ is the number of on-axis radio jets detected, and the beaming factor $f_{\rm b}\simeq 1-\cos\theta_{\rm j}$ is determined by the jet opening angle $\theta_{\rm j}$. SKA's ability to detect off-axis jets reduces sensitivity to assumptions about $\theta_{\rm j}$ and thus improves the fidelity of $f_{\rm j}$ \citep{vanVelzen2021}.

A crucial test of the jet--accretion connection will be coordinated SKA+X-ray campaigns searching for correlations between radio luminosity and accretion disk properties \citep{Mummery2025}, analogous to state transitions in X-ray binaries but scaled to SMBH masses \citep{Merloni2003,Fender2004}. In this framing, late-time radio plateaus or rebrightenings \cite[e.g., AT2020vwl;][]{Goodwin2025} may diagnose accretion-state changes, CNM density shells, or delayed energy injection. With sufficient cadence, SKA's broad frequency coverage will allow these scenarios to be distinguished empirically. Finally, SKA–VLBI will constrain sizes, proper motions, and brightness
temperatures \citep{Shu.1.2026.SKA}; for the closest events, milliarcsecond imaging will reveal or limit apparent superluminal motion and resolve jet geometry. 
The one directly resolved TDE jet to date was detected in Arp 299B, where \citet{Mattila2018} measured its deceleration and subluminal expansion at an average speed of $\sim 0.2 c$. The only other detection of motion in a TDE is the subluminal expansion at $\sim 0.05c$ in WTP14adeqka \citep{Golay2025}.
Phased SKA baselines will substantially improve astrometric accuracy and extend the time baseline over which such events can be monitored, enabling more precise measurements of TDE jet motions, and thereby closing the loop between jet incidence, beaming, and accretion physics.

A robust measurement of the radio-selected jetted fraction as a function of black-hole mass will require $\gtrsim 30$ off-axis or mildly relativistic jets followed with a logarithmically-spaced cadence through the first $\sim 100$\,d. That sample would constrain $f_{\rm j}$ to about 30\% in two mass bins. Most such events should lie in the 10--100\,$\mu$Jy range at $z\sim 0.1$--0.3. Present facilities reach that regime for only a few selected sources and cannot sustain the cadence needed to separate intrinsic evolution from scintillation while securing a nuclear localization.

\subsection{How do SMBHs grow?
}

TDE rates trace how stars are fed into the loss cone of galactic nuclei and therefore inform the long-term growth of SMBHs, especially toward the low-mass end and in nuclear star clusters. The per-galaxy disruption rate depends on black-hole mass, stellar density profiles, and the efficiency of loss-cone refilling, with theory and observations indicating characteristic values of $\sim 10^{-5}$--$10^{-4}\,\mathrm{yr}^{-1}$ \citep{Magorrian1999,Stone2016,Yao2024}, albeit with large, host-dependent dispersion and pronounced enhancements in post-starburst systems \citep{French2020,LawSmith2017}. Whether these intermittent accretion episodes contribute a significant fraction of SMBH mass growth remains debated \citep{LuKumar2018,Dai2021}, because radiative energetics alone underconstrain the mass actually accreted versus expelled in winds or jets.  

SKA will address this by delivering radio calorimetry of outflows, thereby quantifying kinetic energy budgets for both jets and quasi-spherical winds, uncovering obscured or dust-enshrouded TDEs that are faint in optical/UV surveys \citep{Mattila2018}, and providing kinetic-to-radiative energy ratios across host types and black-hole masses. Equipartition analyses of broadband radio SEDs \citep{BarniolDuran2013} will yield the energy of the outflow, $E_{\rm k}$, and its size, $R$, as a function of time for large, uniformly-followed samples. Nearby thermal TDEs (e.g., ASASSN-14li) typically exhibit $E_{\rm k}\sim 10^{48}$--$10^{49}\,$erg \citep{Alexander2016,vanVelzen2016}, while jetted TDEs carry beaming-corrected energies $\sim 10^{50}$--$10^{52}\,$erg \citep{Bloom2011,Zauderer2011}. With SKA sensitivities probing larger volumes, these quantities can be measured for tens to hundreds of events, anchoring the kinetic feedback budget and, via comparison to bolometric outputs, constraining the fraction of fallback mass that is ultimately accreted \citep{Bonnerot2020}. Moreover, sensitivity at low frequencies and late times closes the energy accounting for delayed radio brightenings (e.g., AT2018hyz; \citealt{Cendes2022}), thus constraining total injected energy and coupling to the CNM. By correlating SKA-detected radio TDEs with host morphology and star-formation history \citep{vanVelzen2021}, we can further test predicted rate enhancements in nuclei hosting nuclear star clusters or SMBH binaries \citep{Chen2011,Stone2017}, thereby constraining dynamical feeding channels and their role in SMBH growth over cosmic time.

The key quantity here is the ratio of outflow energy to radiated energy, $E_{\rm k}/E_{\rm rad}$, as a function of host type and black-hole mass. Resolving a dispersion of $\lesssim 0.3$\,dex in each major host class (binned by, e.g., mass, star formation rate) requires roughly fifty events with well-sampled broadband spectra. The constraint is much stronger if the synchrotron peak can still be recovered after it shifts below 1\,GHz at late times, which is feasible only with SKA--Low.

\subsection{What is the black-hole mass and spin distribution revealed by TDEs?}

TDEs naturally select the low-mass end of the SMBH population, $M_\bullet\sim 10^{5}$--$10^{7}\,M_\odot$, a regime where classical stellar/gas dynamical and reverberation methods to determine the black hole masses are challenging \citep{Stone2016, Wevers2019}. Observables such as luminosity, color temperature, and rise/decay timescales provide mass estimates \citep{Mockler2019,Hammerstein2023,Mummery2024}, while in selected cases quasi-periodic oscillations or relativistic precession may offer spin diagnostics \citep{Reis2012,Pasham2019, Pasham2024,Zhang2025}. Rapidly-spinning, higher-mass SMBHs can disrupt stars outside the event horizon, admitting ``visible'' TDEs above the canonical Hills mass and thereby extending the accessible mass range \citep{Kesden2012,Leloudas2016,Gafton2019}. 

SKA observations will add the kinetic dimension to these demographic inferences: radio detections and upper limits, when combined with optical/X-ray data, map radio-loudness and kinetic efficiency as functions of $M_\bullet$ and, where available, spin proxies. In this way, SKA samples will populate and refine the low-mass tail of SMBH--host scaling relations using TDE-selected nuclei, with emerging evidence that TDEs extend the SMBH--bulge relation toward smaller $M_\bullet$ \citep{vanVelzen2021,Wevers2017}. With a sufficiently large sample of jetted TDEs, this would also allow us to investigate the impact of black hole mass and spin on jet launching.

SKA-VLBI will allow for localization of off-nuclear TDE candidates to compact star clusters providing access to the IMBH regime ($M_\bullet\sim 10^{4}$--$10^{5}\,M_\odot$). Such localizations are important, as demonstrated by AT2024tvd, because even off-nuclear TDEs can harbor SMBHs \citep{2025ApJ...985L..48Y}. In terms of probing the black hole mass function down to the IMBH range, white dwarf (WD)-IMBH TDEs offer an excellent avenue of exploration. To date, there are no confirmed WD-IMBH TDEs, however the launch of Einstein Probe has given rise to a number of potential events \citep{OConnor2025, Li2025}. The radio emission associated with IMBHs could span a broad range of luminosity-timescale parameter space. However, it may not be possible for IMBHs to produce jets in TDEs, if they are formed via hierarchical mergers \citep{2002ApJ...576..894M}, as their spin may not be sufficiently high to launch a jet. 
SKA follow-up will test for jet or wind signatures, and thereby help to constrain IMBH demographics in star clusters and dwarf galaxies \citep{Lin2018,Mezcua2017,LiuYQ2025}.

Although spin measurements from TDEs remain rare and model-dependent, a correlation between radio jet power and independent spin indicators (e.g., high-frequency X-ray timing) would test spin-powered jet scenarios \citep{Tchekhovskoy2014,BlandfordZnajek1977}. SKA's role is to supply robust kinetic power measurements across statistically meaningful samples, enabling tests of how SMBH spin couples to magnetic flux and jet efficiency on cosmic scales.

A useful demographic goal is a uniform census of radio-loudness and kinetic efficiency across $M_{\bullet}=10^{5}$--$10^{7}\,M_\odot$, together with a smaller but clean subset of IMBH candidates. That requires of order one hundred radio-detected TDEs across the low-mass SMBH regime and at least ten milliarcsecond-localized IMBH candidates. For the bulk SMBH population the gain is completeness. For the IMBH channel it is secure source identification, because only SKA--VLBI can exclude SMBHs in satellites or disturbed hosts of the AT2024tvd type at the relevant flux densities.

\subsection{What is the accretion history of dormant SMBHs?
}

\begin{figure}
    \centering
    \includegraphics[width=0.5\linewidth]{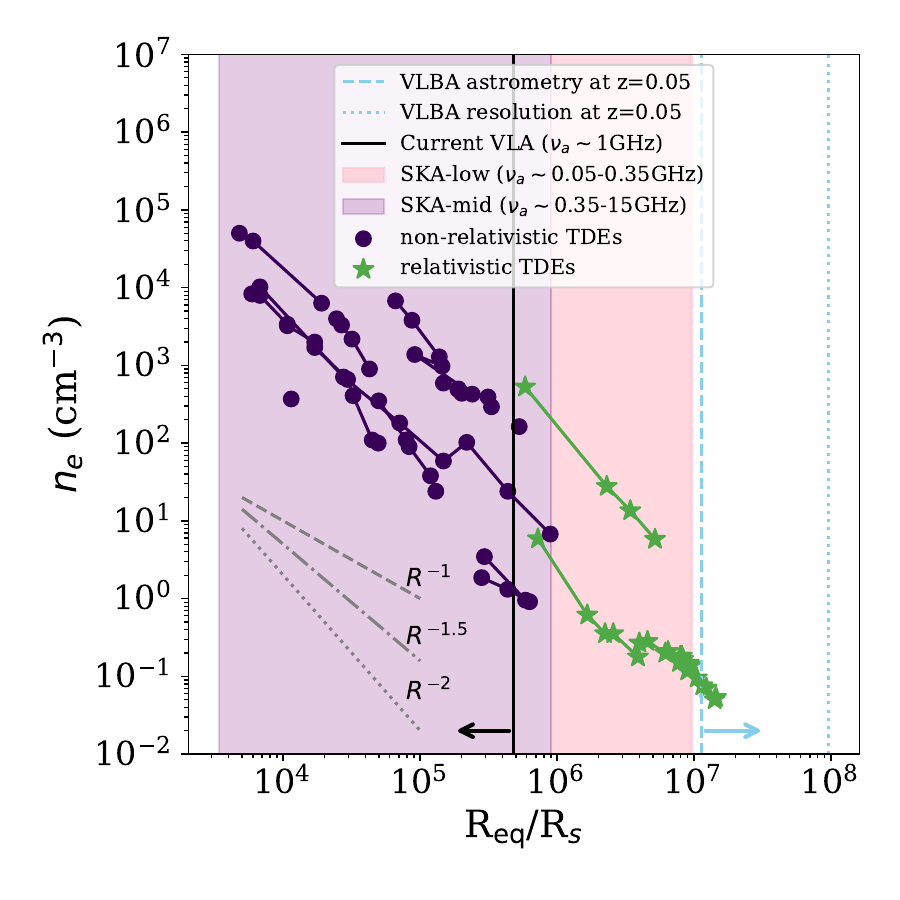}
    \caption{Ambient density $n_{\rm e}$ versus radius inferred from the synchrotron self-absorption break for known TDEs with broadband radio spectra. Current facilities (VLA, ATCA, MeerKAT) reach the SSA break only at $\nu_{\rm p}\gtrsim 1$\,GHz, confining equipartition-derived densities to $r\lesssim 10^{16}$\,cm in all but the brightest relativistic events (shaded region). SKA--Low covers 50--350\,MHz and reaches $\sim$13\,$\mu$Jy\,beam$^{-1}$ in a 1\,hr integration at 150\,MHz. This allows us to detect thermal TDEs out to $D_{L}\lesssim 300$\,Mpc, and extends the region of CNM that can be probed by an order of magnitude. Radio data are drawn from \citet{Alexander2016,Eftekhari2018,Mattila2018,Anderson2020,Cendes2021,Cendes2022,Goodwin2022,Goodwin2023a,Goodwin2023b,Goodwin2024,Christy2024,Rhodes2025}.}
    \label{fig:density_radius}
\end{figure}

As TDE-driven outflows shock the circumnuclear medium (CNM), their radio afterglows encode the ambient density profile $n(r)$ and shock microphysics ($\epsilon_e$, $\epsilon_B$) through the evolving synchrotron SED \citep[e.g.,][]{Chevalier1998,Metzger2016}. Thermal TDEs like ASASSN-14li \citep{Alexander2016,vanVelzen2016,Bright2018} generally indicate non-relativistic outflows expanding into approximately $r^{-2}$ density fields, consistent with wind-shaped CNM environments (Figure \ref{fig:density_radius}). In contrast, late-time flattening or rebrightening in other events such as AT2018hyz, AT2018cqh, and AT2019azh \citep{Cendes2022,Goodwin2022, Yang2025} may reveal strong inhomogeneities, such as clumps, shells, or stratified media, possibly sculpted by past AGN or starburst activity. The broadband sensitivity of the SKA, spanning low to high frequencies, will allow simultaneous tracking of the self-absorption turnover $(\nu_{\rm p},F_{\rm p})$, yielding the temporal evolution of the shock radius $R(t)$ and magnetic field $B(t)$ with minimal modeling assumptions \citep[e.g.][]{BarniolDuran2013,Matsumoto2022}. 

Population-level stacking of long-term TDE light curves with SKA will recover characteristic $n(r)$ profiles across different host-galaxy types and nuclear histories, testing predictions from hydrodynamic simulations of nuclear environments \citep[e.g.,][]{Generozov2017,Mockler2023}. SKA--Low observations are especially powerful at late times, when optically thin emission persists below $\sim$350\,MHz and the blast wave expands to parsec scales, directly probing the density structure of the inner CNM (Figure \ref{fig:density_radius}). Current radio telescopes that operate at these low frequencies are not sensitive enough to detect the fading radio emission from the majority of TDEs at such late times, something only SKA-Low will have the sensitivity to achieve. In addition, high-fidelity polarization imaging and rotation-measure (RM) synthesis \citep{Brentjens2005} will constrain the magnetized plasma surrounding the shocks, enabling spatially resolved estimates of magnetic-field strengths, turbulence, and coherence lengths in galactic nuclei. These measurements will connect present-day outflows to the recent accretion and feedback history of otherwise quiescent SMBHs, providing key diagnostics of how nuclear activity imprints its environment on sub-parsec scales.

This program depends on following $\nu_{\rm p}(t)$ after it moves below 1\,GHz, because that evolution constrains $n(r)$ on scales of 0.1--1\,pc. About twenty events followed with SKA--Low at four to six epochs out to $\sim 1000$\,d would recover the average density profile beyond the Bondi radius. The relevant parameter space is late-time, low-frequency, and sub-mJy, and it remains largely unmeasured so far.

\subsection{What is the population of recoiling or wandering SMBHs/IMBHs?
}

Flares occurring hundreds of parsecs to several kiloparsecs from galactic centers can betray recoiling SMBHs launched by gravitational-wave kicks, SMBHs in merging satellites, or IMBHs bound to dense clusters \citep[e.g.,][]{Lousto2011,Komossa2012,Blecha2016,Bellovary2019,Greene2021}. Recent X-ray and optical discoveries include off-nuclear candidates with TDE-like spectra and light curves (e.g., the IMBH TDE candidate in a star cluster 3XMM~J2150--0551), but robust confirmation requires dust-insensitive energetics and precise astrometry \citep{Lin2018}. 
Another interesting class is luminous fast blue optical transients (LFBOTs), which are suggested to be powered by TDEs involving IMBHs \citep{Gutierrez2024, Inkenhaag2025}. 
The SKA will survey for such off-nuclear transients with accurate positions, follow optical/IR candidates to assess synchrotron counterparts. In addition, SKA--VLBI can refine localizations to milliarcsecond precision to confirm offsets, characterize host environments, and, in exceptional multi-year cases, measure proper motions \citep{Paragi2015}. Even non-detections are informative, placing stringent limits on jet/wind energies and hence on the duty cycle and energetics of wandering IMBHs and recoiling SMBHs \citep{Alexander2020}. A statistically significant SKA-based sample will thus constrain the frequency of displaced massive black holes and their role in galaxy assembly.

The dominant contaminant in this channel is the orphan GRB afterglow, so classification must rely on a purity-controlled combination of observables rather than on any single diagnostic. Candidate off-nuclear TDEs should show a statistically significant displacement from the host nucleus, radio spectral and temporal evolution slower than that of classical long-GRB afterglows, and similarly longer-lasting broadband behaviour consistent with a TDE-driven outflow. Optical and soft-X-ray information, such as color evolution, spectral shapes, and light curves, can further suppress contamination where available. In this chapter we emphasize the role of SKA in discovery, light-curve characterization, and the construction of a clean candidate sample. Milliarcsecond confirmation with phased SKA--VLBI will be required only for a small subset of the most compelling systems, and is discussed in detail by \citet{Shu.1.2026.SKA}. Even a modest sample of well-characterized off-nuclear candidates would place the first meaningful constraints on the incidence of recoiling SMBHs and IMBH-related TDEs.

\subsection{What are the properties of TDE outflows?
}

\begin{figure}
    \centering
    \includegraphics[width=0.5\linewidth]{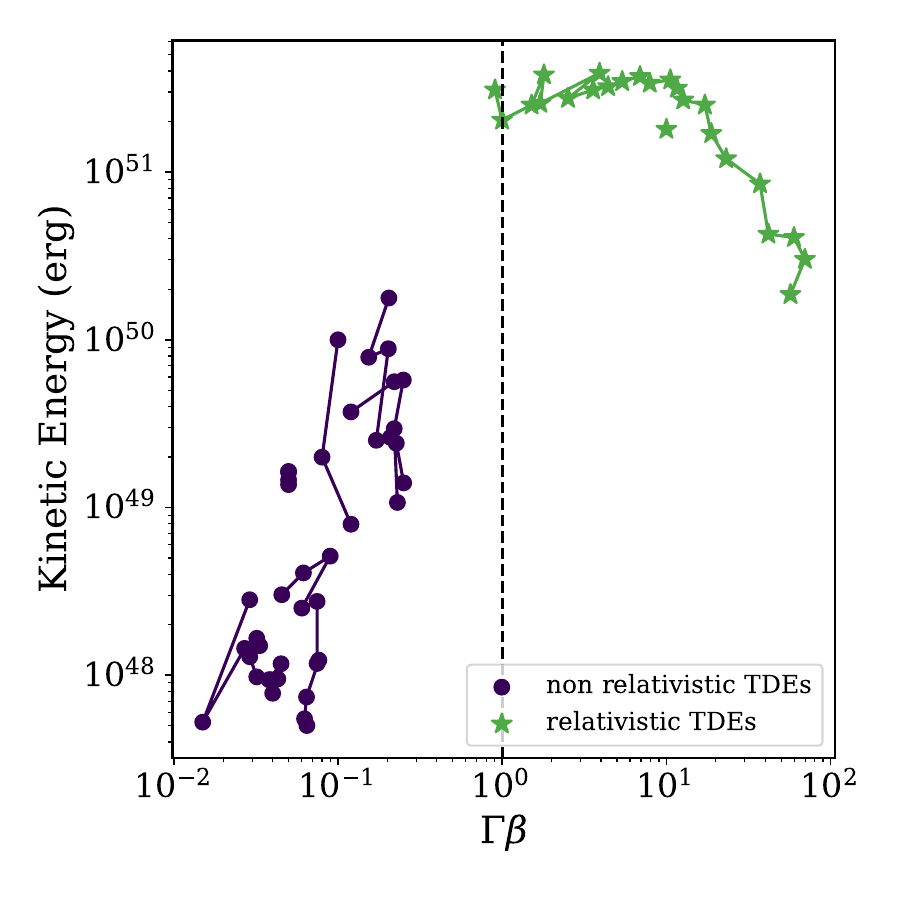}
    \caption{Kinetic energy versus $\Gamma\beta$ for seleceted TDEs with broadband radio spectra, inferred from equipartition analysis. The vertical dashed line separates the relativistic and non-relativistic regimes. Relativistic events cluster at $E_{\rm k}\sim 10^{50}$--$10^{52}$\,erg; non-relativistic outflows span more than two decades in $E_{\rm k}$ for $v\sim 0.05$--$0.3c$. The present sample is dominated by nearby radio-bright events and leaves the low-energy tail sparsely populated. In \AAstar, Band~2 pushes the detectable locus downward by about 1.5\,dex in $E_{\rm k}$ at $z=0.1$ and expands the accessible volume at fixed radio luminosity by a factor of about 30. Data are from \citet{Alexander2016,Eftekhari2018,Mattila2018,Anderson2020,Cendes2021,Cendes2022,Goodwin2022,Goodwin2023a,Goodwin2023b,Goodwin2024}.}
    \label{fig:energy_velocity}
\end{figure}

Broadband radio spectra and their time evolution constrain shock parameters, with equipartition methods yielding characteristic sizes $R$ and kinetic energies $E_{\rm k}$, while temporal slopes measure expansion and deceleration and thereby separate collimated jets from quasi-spherical winds \citep[e.g.,][]{Chevalier1998,BarniolDuran2013}. Observations to date indicate a wide dynamic range: jetted TDEs with $\Gamma\gg 1$ and $E_{\rm k}\sim 10^{51}$--$10^{52}\,$erg \citep{Zauderer2011,Berger2012,Mimica2015,YaoY2023, Rhodes2023}; non-jetted outflows with $v\sim 0.03$--$0.3\,c$ and $E_{\rm k}\sim 10^{48}$--$10^{49}\,$erg \citep[e.g.][]{Alexander2016,Bright2018} (Figure~\ref{fig:energy_velocity}); and late-time rebrightenings that imply refreshed shocks or density structure in the CNM \citep{Cendes2022,Goodwin2022,Cendes2024,Goodwin2025}. With dense, multi-band sampling, the SKA will track $\nu_{\rm p}$ and $F_{\rm p}$ to infer $R \propto t^{m}$ and $B \propto t^{-n}$, use closure relations from forward-shock theory to test microphysical assumptions \citep[analogous to Gamma Ray Burst afterglows;][]{Sari1998,Granot2002}, as applied to AT2022cmc; \citep{Yao2024}), and deploy SKA--VLBI size and proper-motion constraints to quantify collimation and apparent superluminal motion \citep{Yang2016}. Sufficient temporal and frequency sampling will probe different electron energy distributions and populations \citep{Rhodes2025}. When signal-to-noise allows, polarization provides geometry and field-order diagnostics; low polarizations in non-jetted events would favor tangled fields and/or strong depolarization, whereas higher, evolving polarization in jetted events at mm--cm bands would support spine--sheath or shock-compressed field configurations. These measurements motivate clear, falsifiable expectations: within the first $\sim$100\,days, most radio-bright jetted TDEs should show a monotonic decline in $\nu_{\rm p}$ with $F_{\rm p}$ roughly flat or gently decreasing due to ongoing energy injection, while non-jetted events should exhibit coupled declines in both; a minority ($\lesssim 20\%$) of thermal TDEs should display late-time ($>300$\,d) rebrightening indicative of CNM structure or refreshed outflows. In all cases, SKA's cadence and frequency leverage will translate spectral--temporal behavior into quantitative energetics, speeds, collimation, and composition diagnostics for the outflows.

\section{Observational strategy and multi-messenger, multi-wavelength synergies}

Our observing plan combines wide-field, commensal slow-transient surveys with rapid triggered follow-up, targeted late-time campaigns, and SKA--VLBI imaging, while coordinating closely with optical, X-ray, and infrared facilities to maximize multi-wavelength return. A practical program begins with a Band~2 commensal survey at a 3--7\,d cadence. It provides candidate counterparts and an initial nuclear localization. Likely events are revisited a few weeks later with targeted observations, and a subset of the strongest TDE candidates is then re-observed for $\sim0.5$\,hr at about 3, 10, 30, 100, 300 and 1000\,d in Bands~2 and 5, with SKA--Low added once the spectral peak moves below 1\,GHz. This cadence is sufficient to track the evolving synchrotron turnover and recover the physical parameters of the source.

External triggers will come from LSST, ATLAS, Roman, Einstein Probe, XMM-Newton target-of-opportunity programs, and any surviving Swift capability. The SKA response should begin within 24--72\,hr. Early multi-band sampling locates the spectral peak close to discovery, while monitoring over the next few weeks measures the rise and early decline. Hard-X-ray triggers and unusually red optical events are characteristic of jetted TDEs, and hence are natural priorities for prompt Band~5 coverage, where free--free absorption is weaker and newly launched jets are easier to isolate.

The events that constrain energetics and CNM structure most strongly require deeper late-time follow-up. We would therefore schedule 1--2\,hr observations between 100 and 1000\,d for thermal TDEs, delayed radio emitters, and candidate off-axis jets. These epochs recover delayed brightening, anchor the energy budget, and extend the CNM profile to larger radii.

Bright transients (flux densities $\gtrsim 0.3$--0.5~mJy) can be imaged with SKA--VLBI in Band 5a to measure sizes and search for proper motion, quantifying collimation and apparent superluminal speeds \citep[e.g. AT2019dsg;][]{Mohan2022}. For nearby (distance $\lesssim 200$~Mpc) thermal TDEs that peak at 50--200~$\mu$Jy, phased SKA sensitivity enables size measurements or stringent limits after a few months of expansion. Existing VLBI results for Swift~J1644+57 \citep{Yang2016} provide a benchmark, and SKA--VLBI extends these constraints to a broader population with improved astrometric precision.

In terms of survey design, Band~2 (0.95--1.7~GHz) offers excellent sensitivity and a wide field of view for discovery, while Band~5 (4.6--15.4~GHz) reduces free--free absorption and enables early detection of the SSA peak. SKA--Low observations at $\sim$120--200~MHz will probe late-time, optically thin phases when the blast wave has expanded to larger radii. Baselines out to 150~km will deliver angular resolutions of roughly 0.3--1.5~arcsec across Bands~2--5, sufficient to separate nuclear emission from star-formation and to mitigate confusion; direction-dependent calibration in complex fields will preserve image fidelity near bright nuclei. Full-Stokes imaging with per-epoch RM synthesis will constrain magnetized CNM, with even upper limits informing depolarization and Faraday dispersion. Finally, the SKAO's use of tied-array beams and non-imaging voltage-buffer modes will support VLBI applications, while observatory workflows through the Science Regional Centres will ensure timely delivery of calibrated data products for TDE follow-up.

A decisive SKA program for TDEs is intrinsically multi-wavelength and multi-messenger. Optical wide-field surveys, ZTF now and LSST through the 2030s, will deliver clean candidate samples with rise times, colours, and host information that support source selection. Given the uncertain lifetime of Swift, the X-ray program must use a wider set of facilities. Einstein Probe is the near-term wide-field trigger facility for soft-X-ray TDEs and has already uncovered candidate WD--IMBH events \citep{OConnor2025,Li2025}. XMM-Newton remains the best route to deeper spectroscopy and timing while it is operational. Once AA4 is mature, NewAthena should become the high-throughput partner for joint studies of accretion state and radio calorimetry. SVOM provides useful supporting coverage in hard X-rays and the near infrared. Mid-infrared facilities from JWST through Roman will isolate dust-obscured or reprocessing-dominated TDEs that are optically inconspicuous, and SKA radio follow-up then supplies the dust-insensitive kinetic energy term in the energy budget. High-energy triggers add a complementary axis: hard X/\,$\gamma$-ray detections flag candidate jetted events early, and IceCube and KM3Net neutrino alerts motivate rapid SKA campaigns testing hadronic acceleration \citep{Murase2020, Stein2021, Reusch2022}. Milliarcsecond imaging with global VLBI \cite[EVN/VLBA/LBA and African VLBI Network;][]{Bempong-Manful.1.2026.SKA} incorporating phased SKA--VLBI supplies size, geometry, and astrometry; the transient pipelines already developed for supernovae and related radio transients port directly to TDEs.

\section{Expected outcomes, forecasts, and predictions}

In the LSST era the nuclear-transient stream should reach $10^{3}$--$10^{4}$ candidates per year within $z\lesssim 0.5$ \citep{Ivezic2019,vanVelzen2021}. Optical and UV filtering should retain a few $\times 10^{3}$ likely TDEs. In \AAstar, Band~2 commensal visits at 3--7\,d cadence reach 50--100\,$\mu$Jy at 5$\sigma$, and stacking ten visits lowers the threshold to 10--30\,$\mu$Jy, corresponding to $L_{\nu}\gtrsim 10^{28}\,\mathrm{erg\,s^{-1}\,Hz^{-1}}$ at $D_{L}\lesssim 500$\,Mpc. Week-timescale evolution inconsistent with AGN flicker or scintillation, a position consistent with the nucleus to within $0\farcs3$, and a spectrum compatible with synchrotron self-absorption should reduce the sample to 150--300 well-identified nuclear radio TDEs per year from \AAstar\ for a radio-loud fraction $f_{\rm r}\sim 0.1$--0.5.

A realistic baseline program would rely on commensal surveys and external facilities for candidate discovery. Dedicated SKA time would be reserved for intensive follow up of a small, well-selected subset of identified events to constrain jet incidence, CNM structure, and outflow energetics. 
 In practice, 30--50 candidates per year could receive a short Band 5 characterization epoch near optical peak, and the most informative 8--12 events could then be followed at 3, 10, 30, 100, 300, and 1000\,d in Bands 2 and 5, with SKA--Low added only for the nearest late-time emitters. Depending on source brightness and cadence, such a program would require of order 100--200 hr~yr$^{-1}$ and would already define the range of jet and outflow properties, test the classification strategy, and anchor the first SKA measurements of CNM structure. More ambitious programs aimed at precise binning by host class and black-hole mass should be regarded as a later-stage goal, once candidate selection and survey yields are better established. SKA's depth at late times further ensures sensitivity to faint, slowly evolving counterparts that optical surveys lose to dust or host contamination \citep{Alexander2020}.

Off-axis jets should be several times more common than on-axis jetted TDEs; by pushing to low flux densities on week-to-month timescales, SKA will capture their rising afterglows and thereby enable robust beaming corrections and true jet fractions \citep{Metzger2016,Mimica2015,Alexander2020}. Late-time rebrightening appears in a non-negligible subset of thermal TDEs (e.g., AT2018hyz, AT2020vwl, ASASSN-15oi)  \citep{Cendes2022,Goodwin2025,Horesh2021}, likely reflecting refreshed shocks or CNM structure; a dedicated SKA program should detect dozens per year in the LSST epoch, turning such light-curve morphology into environmental diagnostics.

From multi-frequency SEDs and their evolution, SKA will measure $R(t)$, $v_{\rm sh}(t)$, $E_{\rm k}$, and $n(r)$ with typical uncertainties of 20--30\% per well-sampled event, improving when VLBI sizes are available. For nearby thermal TDEs we expect shock velocities $v_{\rm sh}\sim 0.05$--$0.2\,c$ and $E_{\rm k}\sim 10^{48}$--$10^{49}$~erg, while jetted TDEs show beaming-corrected energies $\sim 10^{50}$--$10^{52}$~erg with $\Gamma$ of a few to tens, as inferred for AT2022cmc \citep{Yao2024,Rhodes2023}. SKA--VLBI will deliver mas-scale astrometry and size constraints to quantify collimation and apparent superluminal motion, extending benchmarks set by Swift~J1644+57 and related studies \citep{Paragi2015,Yang2016}. Stacked spectra and light curves across host classes will recover average $n(r)$ profiles and their variance, constraining wind-shaped $r^{-2}$ versus isothermal environments and connecting present-day shocks to prior nuclear activity \citep{Alexander2020,Metzger2016}. Jointly with optical/X-ray mass estimates, SKA radio energetics will extend SMBH--bulge relations to $M_\bullet\sim 10^{5}$--$10^{6}\,M_\odot$, quantify scatter, and test for pseudobulge offsets, while IMBH candidates will be assessed for radio outflows; non-detections will bound $E_{\rm k}$ and duty cycles \citep{vanVelzen2021,Greene2021}.

These programs motivate concrete, testable predictions. First, if spin and magnetic flux dominate jet launching, the radio-loud (or jetted) fraction $f_{\rm j}$ should increase with $M_\bullet$ and with independent spin proxies; combining SKA detections (including off-axis) with LSST/eROSITA counts should constrain $f_{\rm j}$ to $\pm (30$--$50)\%$ within $\sim 5$~years, with trends across mass and host type decisively tested \citep{Ivezic2019,vanVelzen2021}. Second, at least $\sim 10\%$ of thermal TDEs should exhibit late-time ($\gtrsim 300$~d) radio brightening with $\Delta F/F\gtrsim 2$, indicative of CNM density jumps or refreshed shocks; a significantly lower incidence would disfavour such environments \citep{Cendes2022,Goodwin2022}. Third, a non-zero fraction ($\gtrsim 1\%$) of apparently nuclear optical TDEs should prove off-nuclear at mas precision with SKA--VLBI, implying recoiling or wandering SMBHs and IMBHs; a null result after thousands of events would strongly bound these populations \citep{Paragi2015,Lin2018}. Fourth, polarization should be systematically higher (and more variable) in jetted events than in non-jetted outflows, with RM evolution tracing decreasing CNM columns; persistently low polarization would imply strong depolarization or tangled fields, as seen in GRB afterglows \citep{Granot2003}. Finally, if the proposed neutrino--TDE association is real, neutrino-tagged TDEs should show above-average radio luminosities or unusually hard spectra consistent with hadronic acceleration; coordinated SKA campaigns will enable decisive tests \citep{Stein2021, Reusch2022}.

\section{Legacy, Impact, and Summary}

The SKA will move TDE and nuclear-transient studies from sparse, heterogeneous radio follow-up to systematic, well-characterized samples with uniform depth, cadence, and frequency coverage, enabling robust population inferences on jet launching, SMBH demographics, and nuclear feedback. The combination of survey speed, rapid response and deep targeted follow-up of the SKA, together with SKA+VLBI capacity will deliver open, high-value data products, including images, dynamic spectra, polarization cubes, and (for VLBI) calibrated visibilities, curated via the SKA Regional Centres. These archives will power machine-learning classification of nuclear transients and provide a training ground for the next generation of time-domain radio astronomers, while lessons from SKA+VLBI pipelines, which are already recognized as broadly applicable to TDEs, FRBs, and relativistic mergers, will ensure maximal cross-fertilization across transient classes.

With its $\mu$Jy sensitivity, broadband frequency coverage, and flexible observing modes, the SKA will decisively expand and characterize the radio TDE population, including faint or delayed emitters and off-axis jets. Early-time, multi-band radio observations will constrain jet incidence, beaming, and energetics, providing rigorous tests of spin/magnetic flux--driven launching scenarios and their linkage to accretion-state transitions. By tracking broadband spectral and temporal evolution, the SKA will perform CNM tomography, mapping nuclear gas density profiles and magnetization, and thereby connect observed outflows to the recent accretion and feedback history of otherwise dormant SMBHs. At the same time, TDE-selected SMBH samples, extended by the SKA to lower BH masses and to off-nuclear systems, will refine SMBH--host scaling relations and quantify the incidence of IMBHs and recoiling or wandering SMBHs. Finally, coordinated multi-messenger programs with optical surveys (LSST), X-ray missions (Einstein Probe, SVOM), gamma-ray observatories (CTA), and neutrino observatories (IceCube) will test hadronic acceleration scenarios and critically assess any neutrino--TDE connection, stressing the SKA's central role in time-domain, multi-messenger astrophysics.

\section*{Acknowledgements}
MPT acknowledges financial support from the Severo Ochoa grant CEX2021-001131-S and from the Spanish grant PID2023-147883NB-C21, funded by MCIU/AEI/ 10.13039/501100011033, as well as support through ERDF/EU.

\bibliographystyle{abbrvnat-maxbibnames4}
\bibliography{SKA-TDE}

\end{document}

%% file: journal-names.tex
\newcommand{\actaa}{Acta Astron.} 
\newcommand{\araa}{ARA\&A} 
\newcommand{\aar}{A\&ARv} 
\newcommand{\aapr}{A\&ARv} 
\newcommand{\ab}{Astrobiol.} 
\newcommand{\aj}{AJ} 
\newcommand{\apj}{ApJ} 
\newcommand{\apjl}{ApJL} 
\newcommand{\apjs}{ApJSS} 
\newcommand{\ao}{Appl. Opt.} 
\newcommand{\apss}{Astro. \& Space Sci.} 
\newcommand{\aap}{A\&A} 
\newcommand{\aaps}{A\&AS.} 
\newcommand{\baas}{Bull. Am. Astron. Soc.} 
\newcommand{\caa}{Chinese A\&A} 
\newcommand{\cjaa}{Chinese J. A\&A} 
\newcommand{\cqg}{Class. Quantum Gravity} 
\newcommand{\gal}{Galaxies} 
\newcommand{\gca}{Geo. Cosmo. Acta} 
\newcommand{\icarus}{Icarus} 
\newcommand{\jcap}{JCAP} 
\newcommand{\jgr}{J. Geophys. Res.} 
\newcommand{\jgrp}{J. Geophys. Res. Planets} 
\newcommand{\jqsrt}{J. Quant. Spectrosc. Radiat. Transf.} 
\newcommand{\memsai}{Mem. SAIt} 
\newcommand{\mnras}{MNRAS} 
\newcommand{\nat}{Nature} 
\newcommand{\nastro}{Nat. Astron.} 
\newcommand{\ncomms}{Nat. Commun.} 
\newcommand{\nphys}{Nat. Phys.} 
\newcommand{\na}{New Astron.} 
\newcommand{\nar}{New Astron. Rev.} 
\newcommand{\physrep}{Phys. Rep.} 
\newcommand{\pra}{Phys. Rev. A} 
\newcommand{\prb}{Phys. Rev. B} 
\newcommand{\prc}{Phys. Rev. C} 
\newcommand{\prd}{Phys. Rev. D} 
\newcommand{\pre}{Phys. Rev. E} 
\newcommand{\prx}{Phys. Rev. X} 
\newcommand{\prl}{Phys. Rev. Let.} 
\newcommand{\psj}{Planet. Sci. J.} 
\newcommand{\planss}{Planet. Space Sci.} 
\newcommand{\pnas}{Proc. Natl Acad. Sci. USA} 
\newcommand{\procspie}{Proc. SPIE} 
\newcommand{\pasa}{PASA} 
\newcommand{\pasj}{PASJ} 
\newcommand{\pasp}{PASP} 
\newcommand{\rmxaa}{RMXAA} 
\newcommand{\sci}{Science} 
\newcommand{\sciadv}{Sci. Adv.} 
\newcommand{\solphys}{Sol. Phys.} 
\newcommand{\sovast}{Soviet Ast.} 
\newcommand{\ssr}{Space Sci. Rev.} 
\newcommand{\uni}{Universe} 